\documentclass{elsart}

\usepackage{graphicx}

\usepackage{amssymb}

\begin{document}

\begin{frontmatter}



\title{Axion-Like Particles, Cosmic Magnetic Fields and Gamma-Ray Astrophysics}


\author[udine,altro]{Alessandro De Angelis},
\author[udine]{Oriana Mansutti},
\author[pavia]{Marco Roncadelli}

\address[udine]{Universit\`a di Udine, and INFN Trieste, I-33100 Udine, Italy}
\address[altro]{also at INAF Trieste, Italy and at IST Lisboa, Portugal}
\address[pavia]{INFN Pavia, Via A. Bassi 6, I-27100 Pavia, Italy}

\date{26 November 2007}


\begin{abstract}
Axion-Like Particles (ALPs) are predicted by many extensions of the Standard Model and give rise to characteristic dimming and polarization effects in a light beam travelling in a magnetic field.
In this Letter, we demonstrate that photon-ALP mixing in cosmic magnetic fields produces an observable distortion in the energy spectra of distant gamma-ray sources (like AGN) for ranges of the ALP parameters allowed by all available constraints. The resulting effect is expected to show up in the energy band $100 \, {\rm MeV} - 100 \, {\rm GeV}$, and so it can be serched with the upcoming GLAST mission.
\end{abstract}

\end{frontmatter}




\section{Introduction}

It is generally taken for granted that observations yield fair images of astronomical sources, provided sufficient care is exercised. However, environmental effects on the photon beam -- from the source on its way to us -- can mislead the observer, because unexpected effects can be at work.
This happens e.g.\ when dust extinction and reddening become substantial, or when background magnetic fields affect the polarization state of radiation propagating in a cold plasma, thus producing a Faraday rotation.

Remarkably enough, magnetic fields can also give rise to more subtle -- and physically much more interesting -- dimming and polarization effects in a light beam if photons couple to new hypothetical very light particles, to be referred to as Axion-Like Particles (ALPs).
Turning the argument around, detection of nontrivial effects of this sort can be interpreted as observational evidence in favor of an ALP, thereby yielding a crucial piece of information to go beyond the Standard Model.

Our aim is to show that photon-ALP mixing in cosmic magnetic fields can indeed lead to the detection of ALPs in gamma-ray astronomy.
More specifically, we will demonstrate that photon-ALP mixing produces an observable distortion in the energy spectra of gamma-ray sources at cosmological distances -- typically Active Galactic Nuclei (AGN) -- for ranges of the ALP parameters which are allowed by all available constraints.
The resulting effect is expected to show up in the energy band $100 \, {\rm MeV} - 100 \, {\rm GeV}$, and so it can be serched with the upcoming GLAST mission.

\begin{figure}
\centering
\includegraphics[width=.75\textwidth]{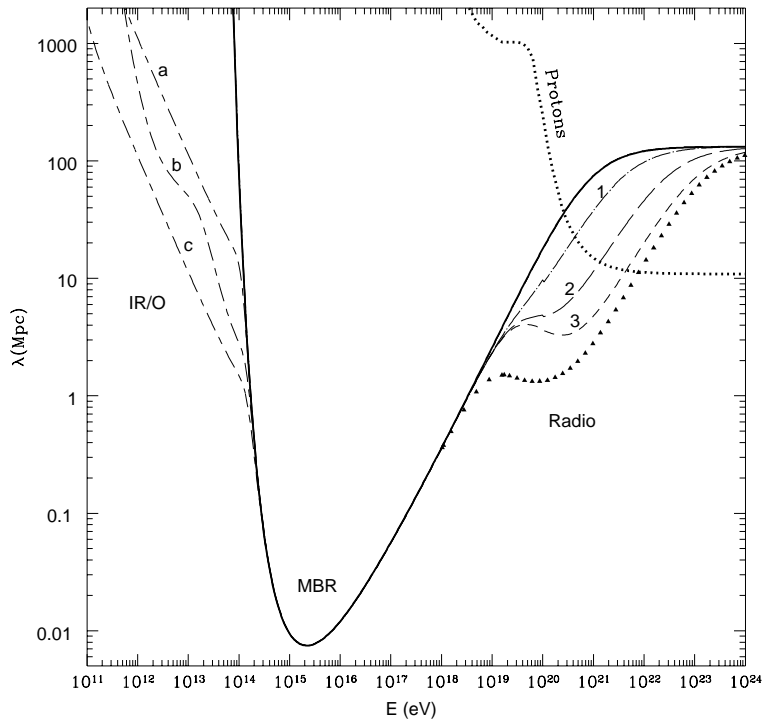}
\caption{\label{CoppiAharonian}
	Pair-production photon mean free path as a function of beam photon energy (from ref.~\cite{CoppiAharonian}).}
\end{figure}

As far as the scope of this Letter is concerned, a further specification is in order.
It is well known that electron-positron pair production in the scattering of beam photons off extragalactic background light (EBL) becomes an important source of opacity whenever the corresponding photon mean free path ${\lambda}_{\gamma}$ is smaller than the source distance $D$.
The energy-dependence of ${\lambda}_{\gamma}$ can be computed within realistic models for EBL and is reported e.g.\ in Fig.~1 (from ref.~\cite{CoppiAharonian}).
Manifestly, the resulting dimming complicates the distortion pattern arising from photon-ALP mixing alone.
In order to achieve a better understanding of the latter mechanism, we find it convenient to presently discard EBL-induced absorption effects, deferring their analysis to a separate publication~\cite{DARMA}.
We will therefore focus throughout on the regime in which ${\lambda}_{\gamma} > D$.
A glance at Fig.~\ref{CoppiAharonian} shows that this situation occurs either for beam-photon energy $E < 10^2 \, {\rm GeV}$ and arbitrary values of $D$, or else for $E > 10^2 \, {\rm GeV}$ provided that the condition ${\lambda}_{\gamma}(E) > D$ is explicitly enforced.

This Letter is structured as follows. Sect.~2 offers a brief overview of the properties of ALPs which are of direct relevance for the subsequent discussion.
Sect.~3 summarizes those features of cosmic magnetic fields in which observable photon-ALP conversion is likely to take place.
Sect.~4 contains the quantitative estimate of the distortion of the energy spectra of AGN arising from photon-ALP mixing.
Finally, we summarize our main conclusions in Sect.~5, where we also compare the proposed mechanism with similar ones recently appeared in the literature.

\section{Photon-ALP Mixing}

The possibility that photon mixing with a light particle alters the physical state of a beam was first recognized in connection with the {\it axion}, the pseudo-Goldstone boson associated with the Peccei-Quinn $U(1)_{\rm PQ}$ global symmetry invented to solve the ``strong CP-problem'' in a natural way~\cite{axion}.
In all viable axion models~\cite{axionreview}, the axion mass is given by \mbox{$m \simeq 0.6 \, ( 10^7 \, {\rm GeV}/f_a ) \, {\rm eV}$}, with $f_a$ denoting the scale at which the $U(1)_{\rm PQ}$ symmetry is spontaneously broken.
The quark-axion Yukawa couplings induce a photon-axion interaction at one-loop (arising from the triangle graph with internal fermion lines), which is described by the effective lagrangian
\begin{equation}
\label{aq5}
{\cal L}_{\phi \gamma} = - \frac{1}{4 M} \, F^{\mu \nu} \,  \tilde F_{\mu \nu} \, \phi =
\frac{1}{M} \, {\bf E} \cdot {\bf B} \, \phi~,
\end{equation}
where $\phi$ stands for the axion field, \mbox{$M \simeq 1.2 \cdot 10^{10} \, k \, ( f_a /10^7 \, {\rm GeV} ) \, {\rm GeV}$} and \mbox{$k\sim1$} is a parameter whose exact value depends on the specific axion model~\cite{cgn} ($M$ is actually independent of the mass of the fermion running in the loop).
Hence, the axion enjoys the characteristic mass-coupling relation
\begin{equation}
\label{eq.m(M)}
m \simeq 0.7 \cdot k \, \left( \frac{10^{10} \, {\rm GeV}}{M} \right) \, 
{\rm eV}~.
\end{equation}

We stress the fact that the lagrangian ${\cal L}_{\phi \gamma}$ naturally arises in a much broader class of realistic models, encompassing four-dimensional extensions of the Standard Model~\cite{masso1}, compactified Kaluza-Klein theories~\cite{kk} and superstring theories~\cite{superstring}.
Accordingly, ${\cal L}_{\phi \gamma}$ is thought to describe ALPs, similar in nature to the axion but with $m$ and $M$ treated as independent parameters~%
\footnote{%
ALPs are supposed to be light enough, and for definiteness one assumes $m < 1 \, {\rm eV}$.
At variance with the axion case, here the existence of ${\cal L}_{\phi \gamma}$ is just regarded as the {\it defining} feature of ALPs without bothering about its origin.%
}.

A straightforward implication of ${\cal L}_{\phi \gamma}$ is that the interaction eigenstates differ from the propagation eigenstates in the presence of a magnetic field ${\bf B}$, thus generating photon-ALP interconversion; the  form of ${\cal L}_{\phi \gamma}$ entails that only photons polarized in the plane containing ${\bf B}$ and the propagation direction mix with ALPs.
As a result, a photon beam traveling in a magnetic field undergoes specific effects.
Exchange of {\it virtual} ALPs affects the polarization state in a selective manner, whereas production of {\it real} ALPs -- occurring for photon energies $E > m$ -- decreases the beam intensity (besides rotating the polarization vector)~\cite{RaffeltStodolsky}.

Several laboratory as well as astrophysical consequences of the photon-ALP mixing have been addressed, in the hope to detect ALPs~\cite{Raffelt1990}.
In particular, the failure to observe ALPs coming from the Sun in the CAST experiment at CERN has set the stark lower bound
\mbox{$M > 1.14 \cdot 10^{10} \, {\rm GeV}$} for \mbox{$m < 0.02 \, {\rm eV}$}~\cite{cast}, which practically coincides with the theoretical bound derived from consideration of globular cluster stars~\cite{Raffelt1990}.
A stronger bound holds for ALPs with \mbox{$m < 10^{- 10} \, {\rm eV}$}: observations of time-lag between opposite-polarization modes in pulsar radio emission~\cite{mohanti} as well as the energetics of supernova 1987a~\cite{raffeltmasso} yield \mbox{$M > 3 \cdot 10^{11} \, {\rm GeV}$}.

We recall that coherent photon-ALP mixing can be regarded as an oscillation process -- much in the same way as it takes place for massive neutrinos -- apart from the fact that  an external~${\bf B}$ field is needed here, due to the spin mismatch.

Suppose for the moment that ${\bf B}$ is homogeneous and let us denote by~$B_T$ its component transverse to the propagation direction of a monochromatic photon beam with energy~$E$.
Then the probability that a photon will convert to an ALP after a distance $x$ reads~\cite{RaffeltStodolsky}
\begin{equation}
\label{a16}
P_{\gamma \to \phi}^{(0)}(x) = {\rm sin}^2 2 \theta \  {\rm sin}^2
\left( \frac{\Delta_{\rm osc} \, x}{2} \right)~,
\end{equation}
where the photon-ALP mixing angle $\theta$ is
\begin{equation}
\label{a16m}
\theta = \frac{1}{2} \, {\rm arcsin} \left( \frac{B_{\rm T}}{M \, {\Delta}_{\rm osc}} \right)
\end{equation}
and the oscillation wavenumber reads
\begin{equation}
\label{a17}
{\Delta}_{\rm osc} = 
\left[\left( \frac{m^2 - {\omega}_{\rm pl}^2}{2 E} \right)^2 + 
\left( \frac{B_{\rm T}}{M} \right)^2 \right]^{1/2}~,
\end{equation}
so that the  oscillation length is  \mbox{$L_{\rm osc} = 2 \pi / {\Delta}_{\rm
osc}$}~%
\footnote{%
Since we are dealing with weak magnetic fields, their contribution to the vacuum refractive index is negligible~\cite{C}.%
}.
In fact, eq.~(\ref{a17}) pertains to the situation in which the beam propagates in a magnetized cold plasma, which gives rise to an effective photon mass set by the plasma frequency \mbox{${\omega}_{\rm pl} = \sqrt{4 \pi \alpha n_e/m_e}$} $\simeq$ \mbox{$3.69 \cdot 10^{- 11} \, \sqrt{n_e /{\rm cm}^{- 3}} \, {\rm eV}$}, where~$n_e$ is the electron density ($m_e$~denotes the electron mass).

\begin{figure}
\centering                                            \hspace*{-17.5mm}
\includegraphics[width=.56\textwidth]{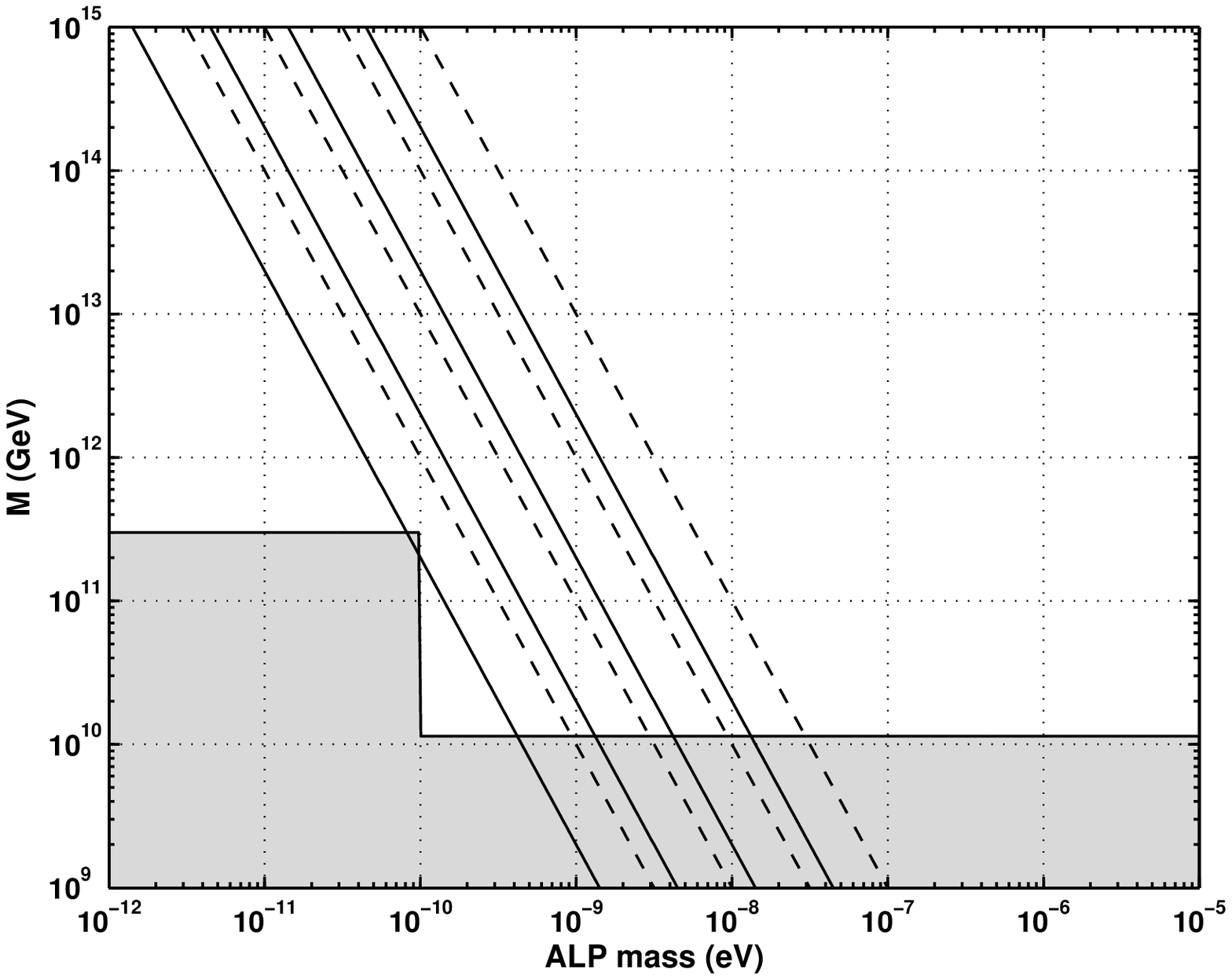} \hspace*{- 7.5mm}
\includegraphics[width=.56\textwidth]{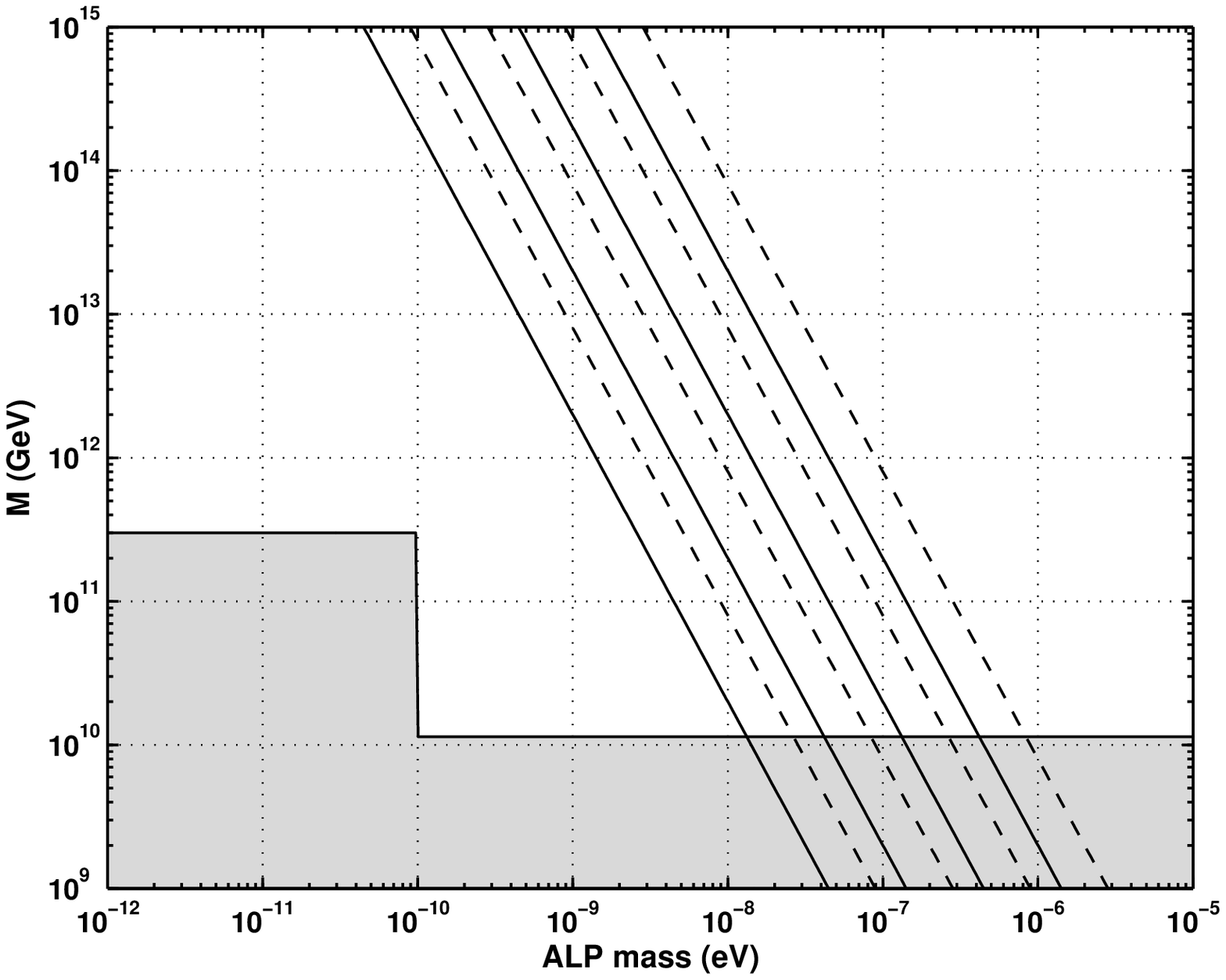} \hspace*{-17.5mm}
\caption{\label{fig:omegaStar}
	Left panel: values of the pair $(m,M)$ which determine the critical energy $E_*$~=~1~GeV, 10~GeV, 100~GeV and 1~TeV (from left to right) for a magnetic field strength of $B = 1\cdot10^{-9}$~G (solid line) and $B = 5\cdot10^{-9}$~G (dotted line) and a plasma frequency $\omega_{\rm pl} \sim 10^{-14}$~eV. The gray region represents the values excluded by astrophysical arguments and by the CAST experiment.
\newline
	Right panel: same as left panel, but with $B = 1\cdot10^{-6}$~G (solid line) and $B = 4\cdot10^{-6}$~G (dotted line) and a plasma frequency $\omega_{\rm pl} \sim 10^{-12}$~eV.}
\end{figure}

A deeper insight into the physics of photon-ALP oscillations can be gained by introducing the critical energy
\begin{equation}
{E}_* \equiv
\frac{|m^2 - {\omega}_{\rm pl}^2| M}{2 B_T} 
\simeq  0.26 \cdot 10^{10} 
\frac{| m^2 - {\omega}_{\rm pl}^2|}{(10^{-10}{\rm eV})^2}
\left( \frac{10^{-9}{\rm G}}{B_T} \right)
\left( \frac{M}{10^{10}{\rm GeV}} \right)
{\rm eV}
\end{equation}
(see Fig.~\ref{fig:omegaStar}).
For further needs, we note that eqs.~(\ref{a16}) and (\ref{a17}) can be rewritten as
\begin{equation}
\label{a16x}
P_{\gamma \to \phi}^{(0)}(x) = \frac{1}{1 + \left( \frac{E_*}{E} \right)^2} \,
{\rm sin}^2\left( \frac{B_T}{M} \, \left[ 1 + \left( \frac{E_*}{E} \right)^2\right]^{1/2} \frac{x}{2} \right)
\end{equation}
and
\begin{equation}
\label{a17y}
{\Delta}_{\rm osc} = \left( \frac{B_T}{M} \right) \left[ 1 + \left( \frac{E_*}{E}
\right)^2\right]^{1/2}~,
\end{equation}
respectively.
Accordingly, the situation can be schematized as follows.

{\it Strong-mixing regime} -- In the high-energy limit \mbox{$E \gg {E}_*$}, we immediately have \mbox{${\Delta}_{\rm osc} \simeq B_T/M$}, the photon-ALP mixing is {maximal} ($\theta \simeq \pi/4$) and the conversion probability becomes energy-independent.

{\it Weak-mixing regime} -- In the opposite low-energy limit \mbox{$E \ll {E}_*$}, we get \mbox{${\Delta}_{\rm osc} \simeq$} \mbox{$\simeq |m^2 - {\omega}_{\rm pl}^2|/ 2 {E}$}.
The mixing is small ($\theta \ll 1$), photon-ALP oscillations become dispersive -- since now both the mixing angle and the oscillation length are energy-dependent -- and their amplitude gets reduced by the factor $({E}/{E}_*)^2$.

In either case, the simpler behavior \mbox{$P_{\gamma \to \phi}^{(0)}(x) \simeq (B_T \, x/2 M)^2$} emerges for an oscillation length \mbox{$L_{\rm osc} \gg x$}.

\section{Cosmic magnetic fields}

As already stressed, photon-ALP conversion requires the presence of a magnetic field playing the role of a catalyst.
Below, we consider those cosmic magnetic fields which are likely to affect in a substantial manner the physical state of a photon beam from a distant AGN (within the ALP scenario outlined in Sect.~2).

Generally speaking, the origin and structure of magnetic fields in the Universe is still unknown.
A possibility is that very small magnetic fields present in the early Universe were subsequently amplified by the process of structure formation~\cite{dolag}.
An alternative option is that magnetic fields have been generated in the low-redshift Universe by energetic quasar outflows~\cite{furlanetto}.
Finally, it has been suggested that the seeds of extragalactic magnetic fields originated from the so-called Biermann battery effect~\cite{biermann}, namely from electric currents driven by merger shocks during structure formation processes.
Presumably, all these effects can take place, even if it is presently impossible to establish their relative importance.

Observations show that cosmic magnetic fields have a complicated morphology, which evidently reflects both the pattern of baryonic structure formation and its subsequent evolutionary history~\cite{magfields}.
In spite of the fact that they come in a wide variety of configurations and strengths, many of them cannot be considered as uniform over the typical distances traveled by photons from cosmological sources.
As a consequence, one cannot evaluate the photon-ALP transition probability by blind application of eq.~(\ref{a16}).

It turns out that -- at least to a first approximation -- one can assume that nonuniform cosmic magnetic fields ${\bf B}$ have a cellular structure (more about this, later).
That is, ${\bf B}$ is supposed to be constant over a domain of size $L_{\rm dom}$ equal to its coherence length, with ${\bf B}$ randomly changing its direction from one domain to another but keeping approximately the same strength.
Over distances \mbox{$D \gg L_{\rm dom}$}, the actual conversion probability $P_{\gamma \to \phi}(D)$ arises as the incoherent average of \mbox{$P_{\gamma \to \phi}^{(0)}(L_{\rm dom})$} over the \mbox{$N \simeq (D/ L_{\rm dom})$} domains crossed by the beam.
One finds~\cite{grossman}
\begin{equation}
\label{h1}
P_{\gamma \to \phi}(D) = \frac{1}{3} \left[ 1 - {\rm{exp}} \left( - \frac{3}{2} \,
\frac{D}{L_{\rm dom}} \, P_{\gamma \to \phi}^{(0)}(L_{\rm dom})  \, \right) \right]~,
\end{equation}
which can be approximated as \mbox{$P_{\gamma \to \phi}(D) \simeq 0.5 \cdot N \, P_{\gamma \to \phi}^{(0)}(L_{\rm dom})$} for $N$ and $L_{\rm dom}$ such that \mbox{$N \, P_{\gamma \to \phi}^{(0)}(L_{\rm dom}) \ll 1$}.
Alternatively, $P_{\gamma \to \phi}(D)$ saturates in the limit
\mbox{$N \, P_{\gamma \to \phi}^{(0)}(L_{\rm dom}) \gg 1$}, so that on average {one-third} of the photons become ALPs.

\subsection{Large-scale magnetic fields} 

So far, observations have failed to detect the existence of magnetic fields over cosmological scales and only upper limits are available on their strength and coherence length. 
Typically, one gets $B < 10^{- 9} - 10^{- 8} \, {\rm G}$ over Megaparsec scales~\cite{magfields,blasi}. 

In the lack of any reliable information, we will carry out our analysis for
\mbox{$B = 1\cdot10^{-9}$~G} and $B = 5\cdot10^{-9}$~G, and
\mbox{$L_{\rm dom} \simeq 1 \, {\rm Mpc}$}, which are close to existing upper limits but consistent with them.

Still, it is interesting to notice that our preferred values are suggested by a simple heuristic argument~\cite{loeb}.
Observations yield $B > 10^{- 7} \, {\rm G}$ in collapsed baryonic structures with  overdensity $\delta \sim 10^3$.
Flux conservation during gravitational collapse (adiabatic compression) entails $B \sim {\delta}^{2/3}$.
So, we get $B > 10^{- 9} \, {\rm G}$ in the intergalactic medium.
Moreover, in the quasar outflow model the cellular structure of the magnetic fields emerges naturally, with a coherence length of a Megaparsec scale.

Besides from magnetic fields, the physical state of the beam propagating over cosmological distances is also affected by the presence of a cold plasma in intergalactic space.
The absence of the Gunn-Peterson effect is usually taken as an evidence that the intergalactic medium is ionized with \mbox{$n_e \simeq 10^{-7} \, {\rm cm}^{-3}$}~\cite{peebles}, resulting in the plasma frequency \mbox{${\omega}_{\rm pl} \simeq 1.17 \cdot 10^{-14} \, {\rm eV}$}, in agreement with the WMAP upper bound \mbox{$n_e < 2.7 \cdot 10^{- 7}  \, {\rm cm}^{-3}$} on the baryon density~\cite{wmap}.

\subsection{Intracluster magnetic fields} 

A better situation concerns clusters of galaxies. Indeed, observations have shown that the presence of magnetic fields with average strength $B \simeq 10^{- 6} \, {\rm G}$ is a typical feature of the intracluster region.
Somewhat stronger values are detected in the cores of regular clusters.
Even more remarkable is the fact that observations are able to yield information about the associated coherence length, which turns out to be of the order of $10$~kpc~\cite{carilli}.
A cellular structure for the intracluster magnetic field is usually assumed, with domain size \mbox{$L_{\rm dom} \simeq 10$~kpc}. 

Just as in the previous case, plasma effects are expected to show up when the beam crosses a cluster.
Specifically, the electron density of the intracluster medium is \mbox{$n_e \simeq 1.0 \cdot 10^{-3} \, {\rm cm}^{-3}$}~\cite{Dolag}, which yields a plasma frequency \mbox{${\omega}_{\rm pl} \simeq 1.2 \cdot 10^{-12} \, {\rm eV}$}.

\subsection{Galactic magnetic fields} 

Observations over the last three decades have led to a rather detailed picture of the magnetic field in the Milky Way.
Perhaps, the most important feature of the Galactic magnetic field is that it consists of two components.

{\it Regular component} --
Measurements of Faraday rotation based on pulsar observations have shown that this component is parallel to the Galactic plane.
Its strength varies between $B \simeq 2 \cdot 10^{- 6} \, {\rm G}$ in the Solar neighbourhood and $B \simeq 4 \cdot 10^{- 6} \, {\rm G}$ at 3 kpc from the centre~\cite{han}.
Moreover, the associated coherence length is of the order of 10 kpc.

{\it Turbulent component} --
Over much smaller scales, the dominant Galactic magnetic field appears to be stochastic, with a Kolmogorov spectrum $\alpha = 5/3$~\cite{ferriere}.
In practice, this component can be described by a cellular structure, with strength $B \simeq 1 \cdot 10^{- 6} \, {\rm G}$ and domain size {$L_{\rm dom} \simeq 10^{- 2} \, {\rm pc}$}.

Inside the Milky Way disk the electron density is \mbox{$n_e \simeq 1.1 \cdot 10^{-2} \, {\rm cm}^{-3}$}~\cite{Digel}, which gives a plasma frequency \mbox{${\omega}_{\rm pl} \simeq 4.1 \cdot 10^{-12} \, {\rm eV}$}.

\section{Spectral distortion}

Our proposal concerns the {\it distortion} of the energy spectra of extragalactic sources like AGN as induced by photon-ALP conversion in intervening magnetic fields (of the kind discussed in Sect.~3).
Presently, the source emission spectrum $dN/dE$ gets modified along the line-of-sight in such a way that at the observer position it becomes $dN/dE$ times the total photon survival probability $P_{\gamma \to \gamma}(D)$.
Because $P_{\gamma \to \gamma}(D) = 1 - P_{\gamma \to \phi}(D)$, we see that the {\it observed} spectral distortion is just $dN/dE \cdot P_{\gamma \to \phi}(D)$.
That is to say, the emission spectrum merely gets distorted in proportion to the photon-ALP conversion probability {\it regardless} of the actual spectral shape.
We stress that this circumstance greatly simplifies our analysis, since it dispenses us from committing ourself with a specific source spectrum.

Owing to eq.~(\ref{h1}), the size of the observed spectral distortion increases with $D$.
Yet, a larger $D$ both makes the source fainter and enhances the EBL-induced absorption at high energies.
In the analysis to follow, we will adopt for definiteness the realistic values $D = 200 \, {\rm Mpc}$ and $D = 500 \, {\rm Mpc}$ whenever necessary.

It goes without saying that the spectral energy distortion has to be measured well enough in order to disentangle the effect in question from other uncertainties.
It should also be kept in mind that a large part of the error in very-high-energy  gamma-ray detectors is correlated~\cite{crosscali}, so that the ratio between the yields in two different energy points can be measured with a relative uncertainty of order 10\% in a Imaging Atmosphreric Cherenkov Telescope and well below 10\% in GLAST.
Thus, a spectral distortion larger than 10\% will be regarded as {\it observable} throughout the subsequent discussion.

We proceed to investigate the behavior of the photon-ALP conversion probability as a function of $E_*/E$.
Over a single magnetic domain, $P_{\gamma \to \phi}^{(0)}(L_{\rm dom})$ is computed from eq.~(\ref{a16x}) and is shown in the left panels of Figs.~\ref{fig:P_omStar_extra_200} and \ref{fig:P_omStar_intra} for suitable values of the magnetic field.
Over the whole distance $D = N \, L_{\rm dom}$, $P_{\gamma \to \phi}(D)$ is given by eq.~(\ref{h1}) and is similarly illustrated in Fig.~\ref{fig:P_omStar_extra_500} and in the right panels of Figs.~\ref{fig:P_omStar_extra_200} and \ref{fig:P_omStar_intra}.
We see that the conversion efficiency increases with energy as long as $E <{E}_*$, while becomes maximal for $E > {E}_*$.
A characteristic feature shows up due to the drastic change in the yield when comparing energies {\it above} and {\it below} $E_*$ by two-three orders of magnitude
-- this is indeed the {\it signature} of the effect we are looking for~%
\footnote{%
In agreement with the discussion in Sect.~2, an oscillatory pattern is present for $E<E_*$, until it becomes unobservable at sufficiently low energy.%
}.

We identify the energy band between 100~MeV and 100~GeV as the best compromise between the detector sensitivity and the lack of EBL-induced absorption.
It is easy to check that condition ${\lambda}_{\gamma}(E) > D$ is presently met.
The energy band in question is almost completely unexplored at present, but it will soon become accessible with the GLAST satellite.

\begin{figure}[bt]
\begin{tabular}{@{}c@{}c@{}}                               \hspace{-5mm}
\includegraphics[width=.56\textwidth]{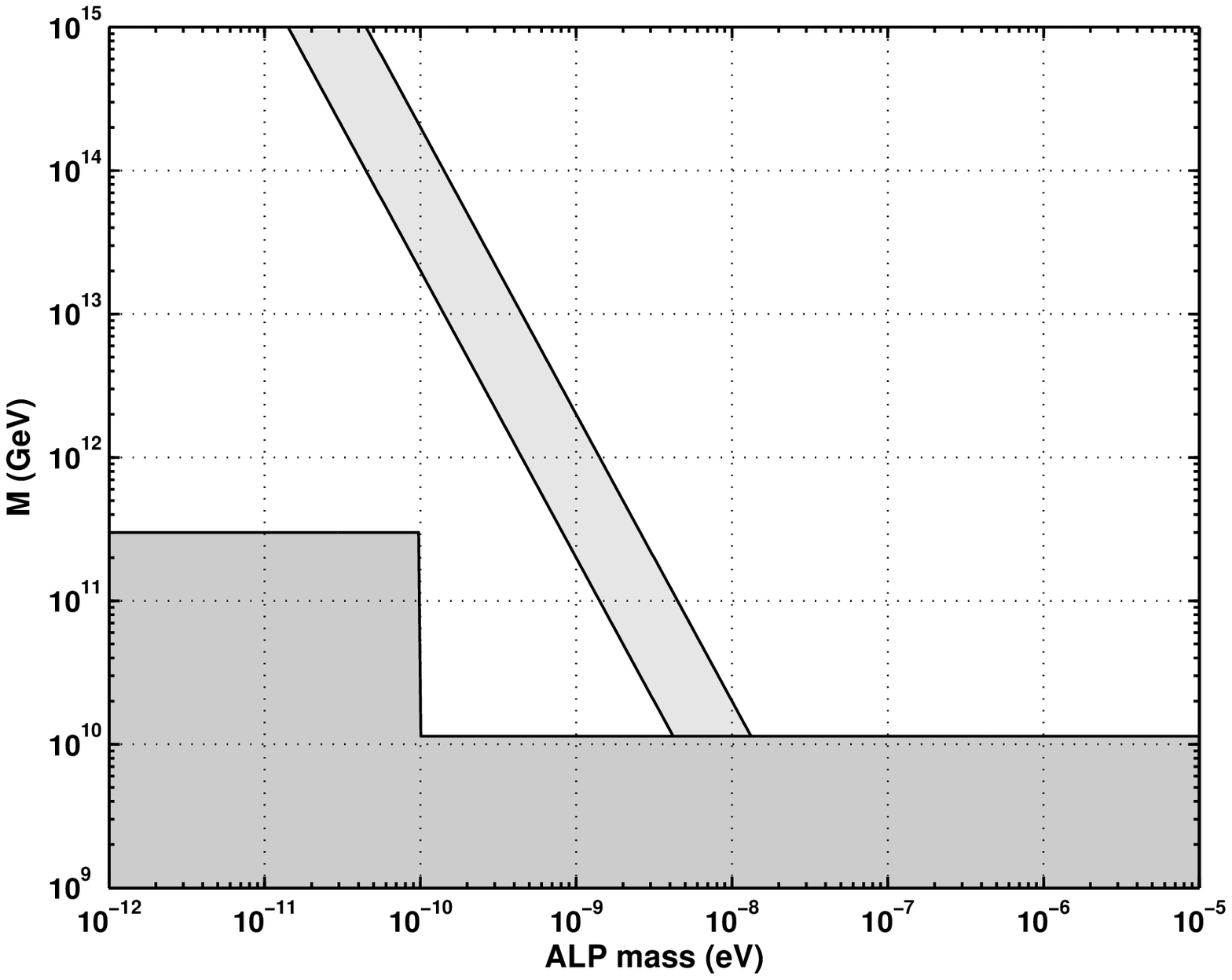}  \hspace{-2.5mm}
&                                                          \hspace{-5mm}
\includegraphics[width=.56\textwidth]{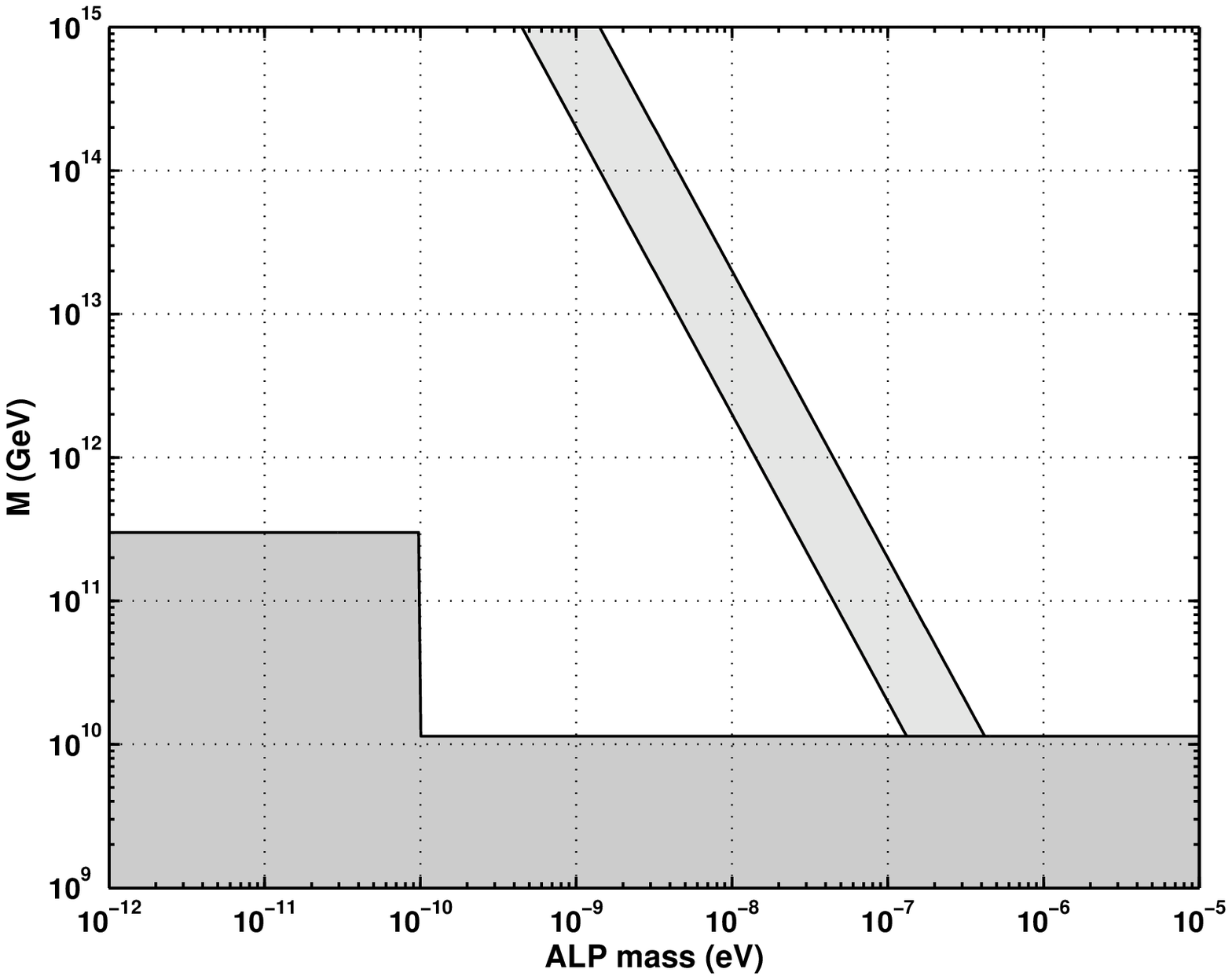}
\\                                                         \hspace{-5mm}
\includegraphics[width=.56\textwidth]{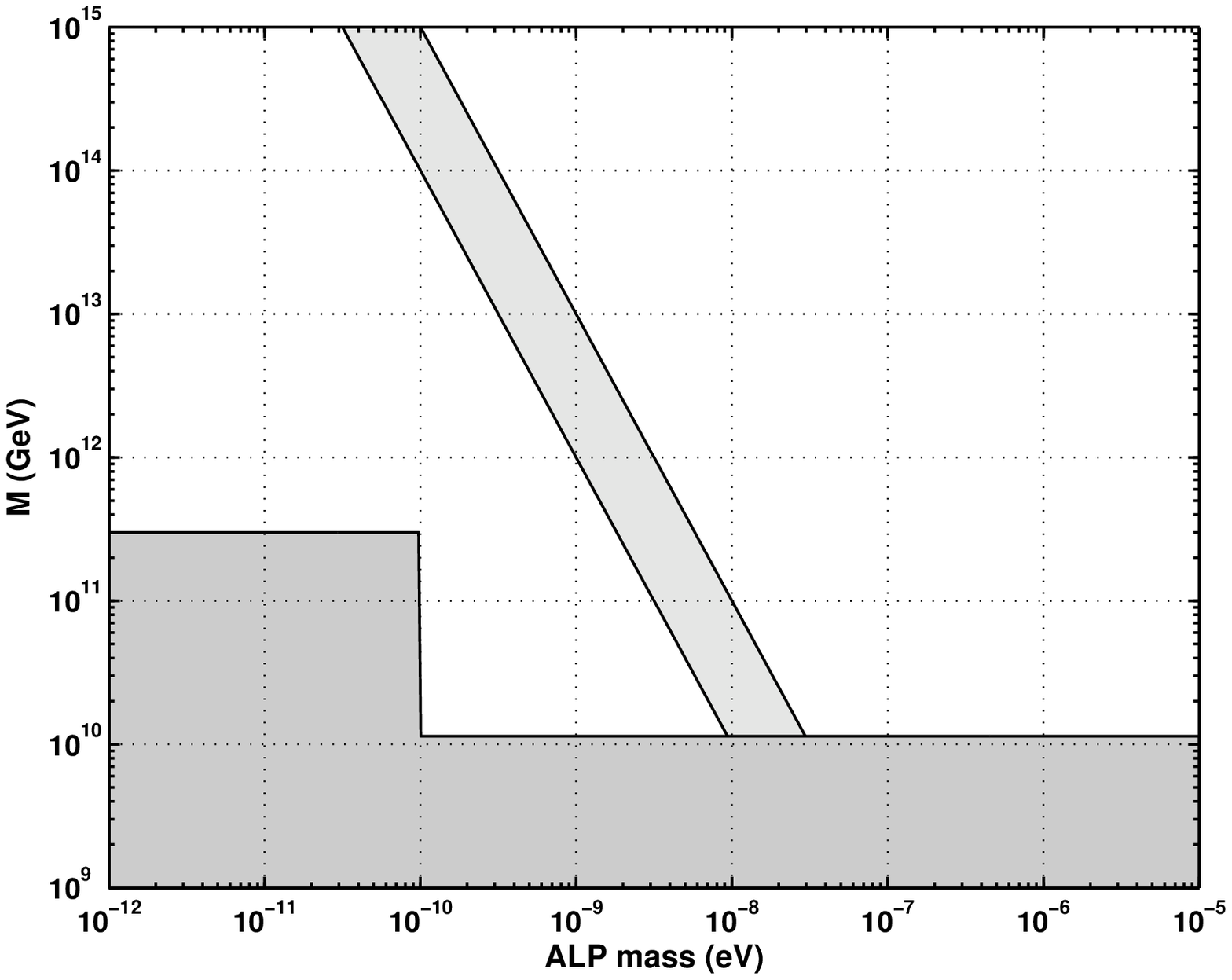} \hspace{-2.5mm}
&                                                          \hspace{-5mm}
\includegraphics[width=.56\textwidth]{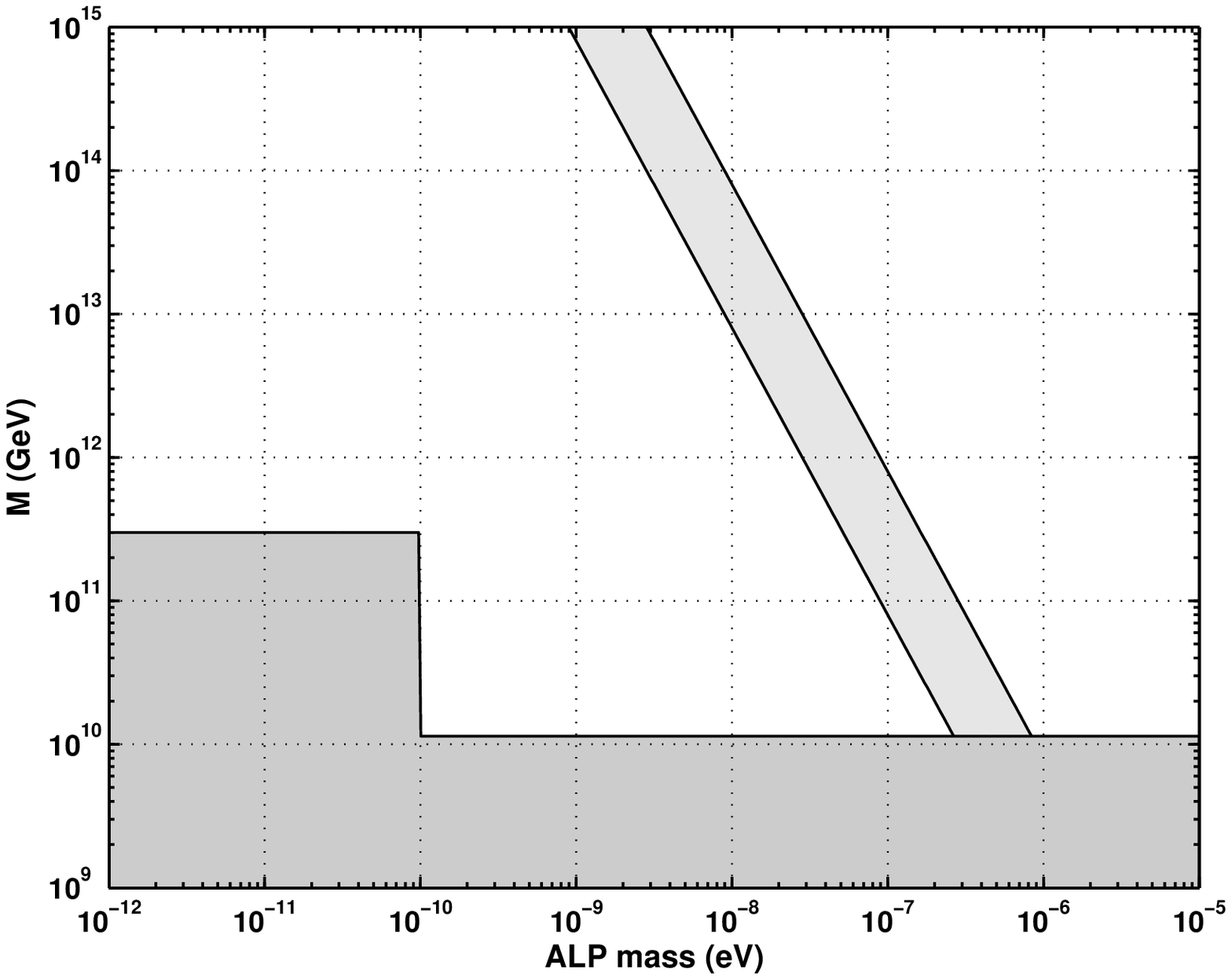}
\end{tabular}
\caption{\label{fig:omegaStar_bands}
	Left panels: region in the $(m,M)$ space which determine the critical energy $E_*$ between 100~GeV and 1~TeV for a magnetic field strength of \mbox{$B = 1\cdot10^{-9}$~G} (upper plot) and $B = 5\cdot10^{-9}$~G (lower plot) and a plasma frequency $\omega_{\rm pl} \sim 10^{-14}$~eV. The dark gray region represents the values excluded by astrophysical arguments and by the CAST experiment.
	\newline
	Right panels: same as left panels with however $B = 1\cdot10^{-6}$~G (upper plot) and $B = 4\cdot10^{-6}$~G (lower plot) and a plasma frequency $\omega_{\rm pl} \sim 10^{-12}$~eV.}
\end{figure}

According to the foregoing discussion, observable effects can be detected in such an energy band provided the critical energy $E_*$ lies {\it just} above its upper edge, namely for $E_* \sim 10^2 \, {\rm GeV} - 1 \, {\rm TeV}$.
The constraints implied by the latter condition on the parameters $m$ and $M$ are reported in Fig.~\ref{fig:omegaStar_bands} for suitable values of the magnetic field.
Correspondingly, we find that in the allowed region of the $(m,M)$ space the less stringent CAST bound on $M$ applies and that plasma effects are unimportant.

Below, we evaluate the spectral energy distortion for a distant AGN as produced by photon-ALP conversion occurring in the three magnetic environments considered in Sect.~3.
Clearly, all we have to do is to evaluate $P_{\gamma \to \phi}(D)$. {\it Observable} effects get singled out by the requirement $P_{\gamma \to \phi}(D) > 0.1$.

\subsection{Large-scale contribution}

\begin{figure}[t]
\begin{tabular}{@{}c@{}c@{}}                               \hspace{-5mm}
\includegraphics[width=.56\textwidth]{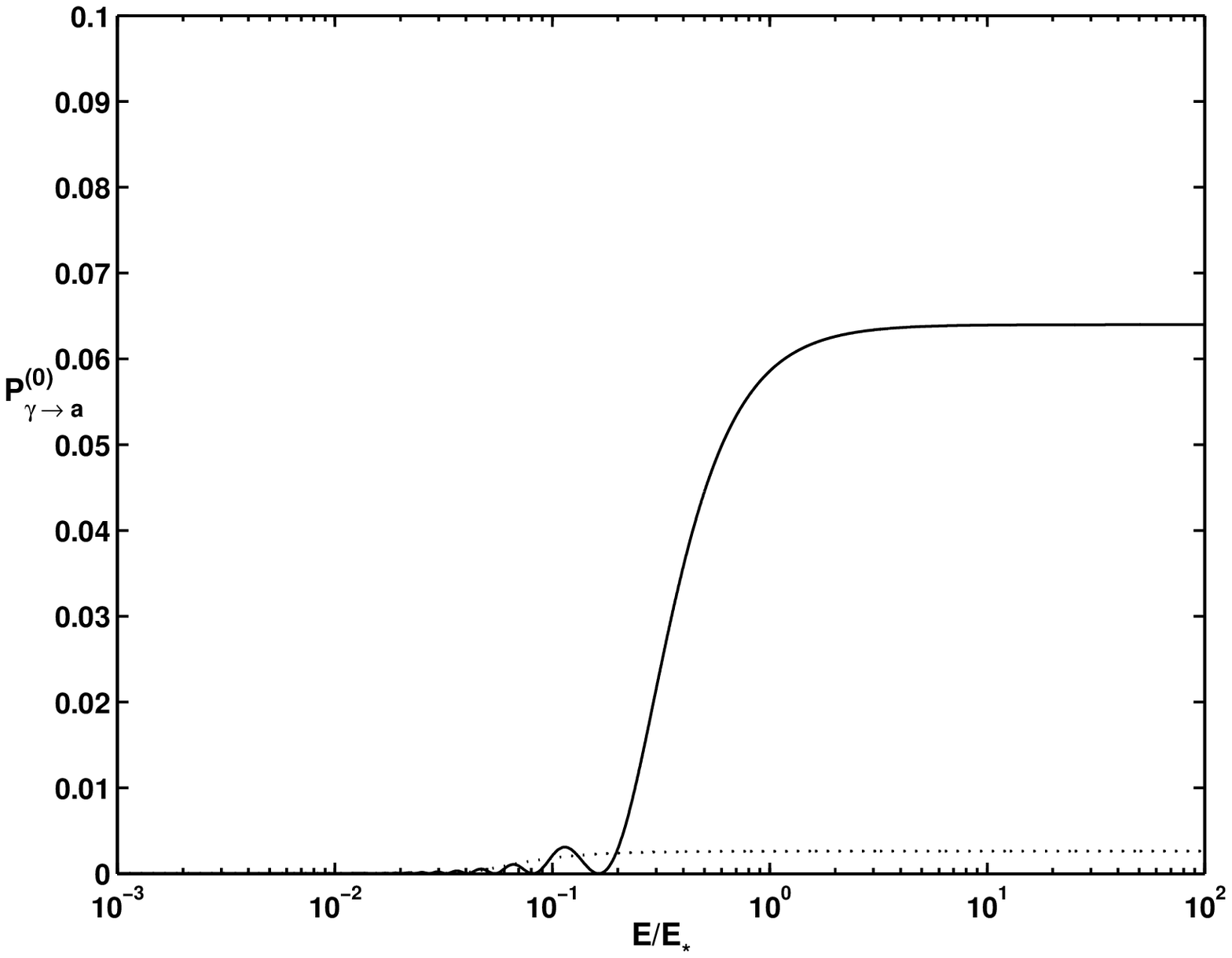} \hspace{-2.5mm}
&                                                          \hspace{-5mm}
\includegraphics[width=.56\textwidth]{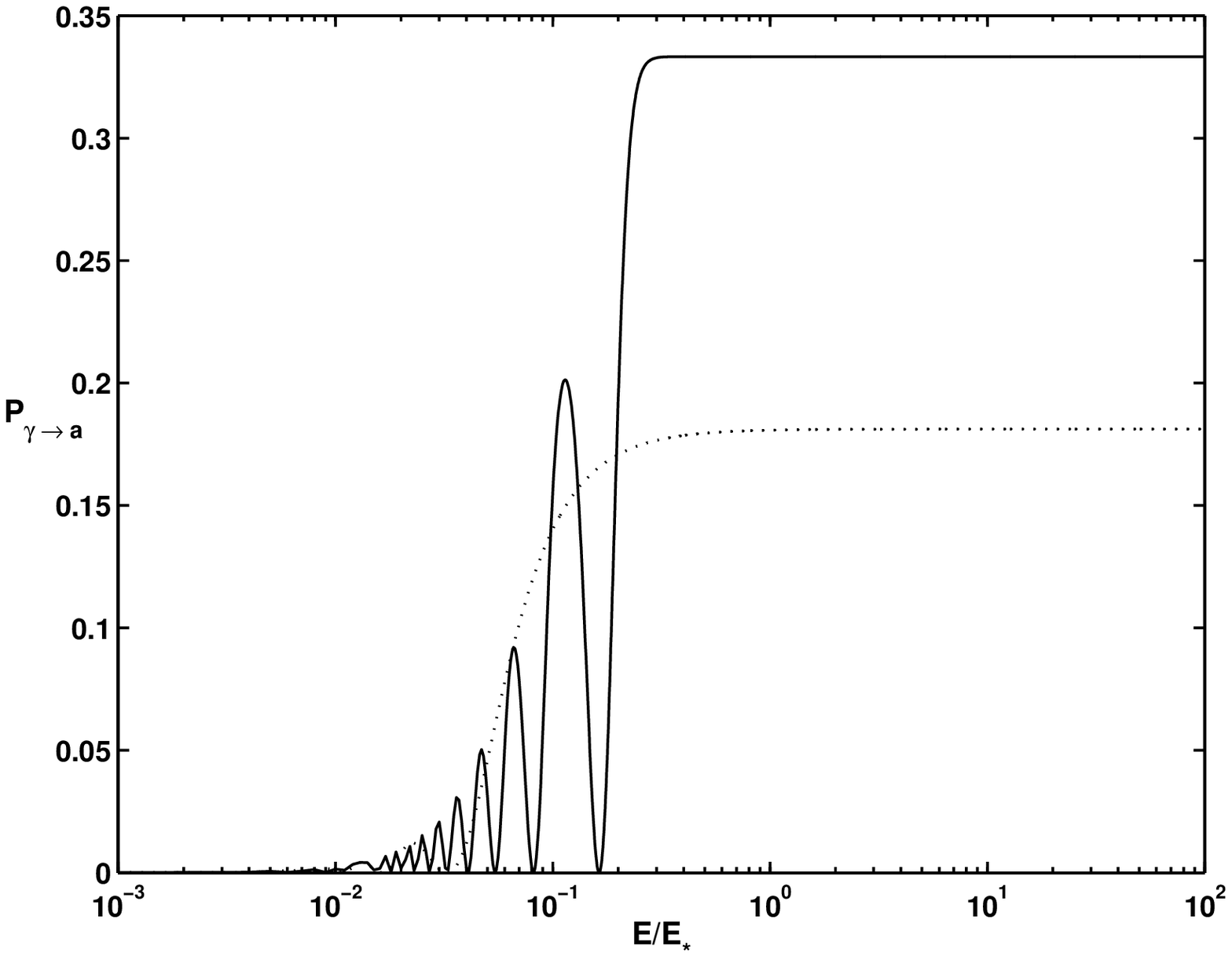}
\\                                                         \hspace{-5mm}
\includegraphics[width=.56\textwidth]{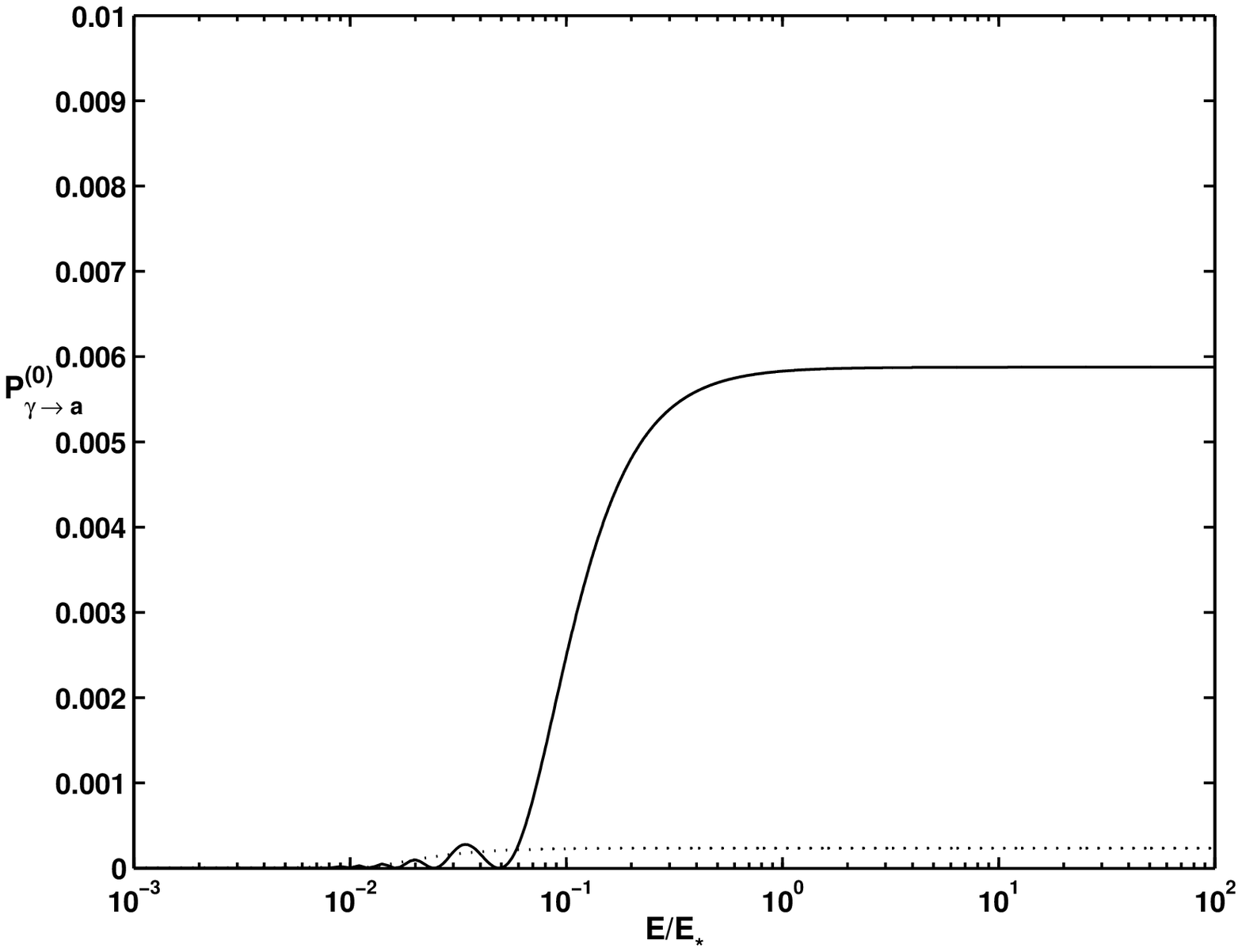} \hspace{-2.5mm}
&                                                          \hspace{-5mm}
\includegraphics[width=.56\textwidth]{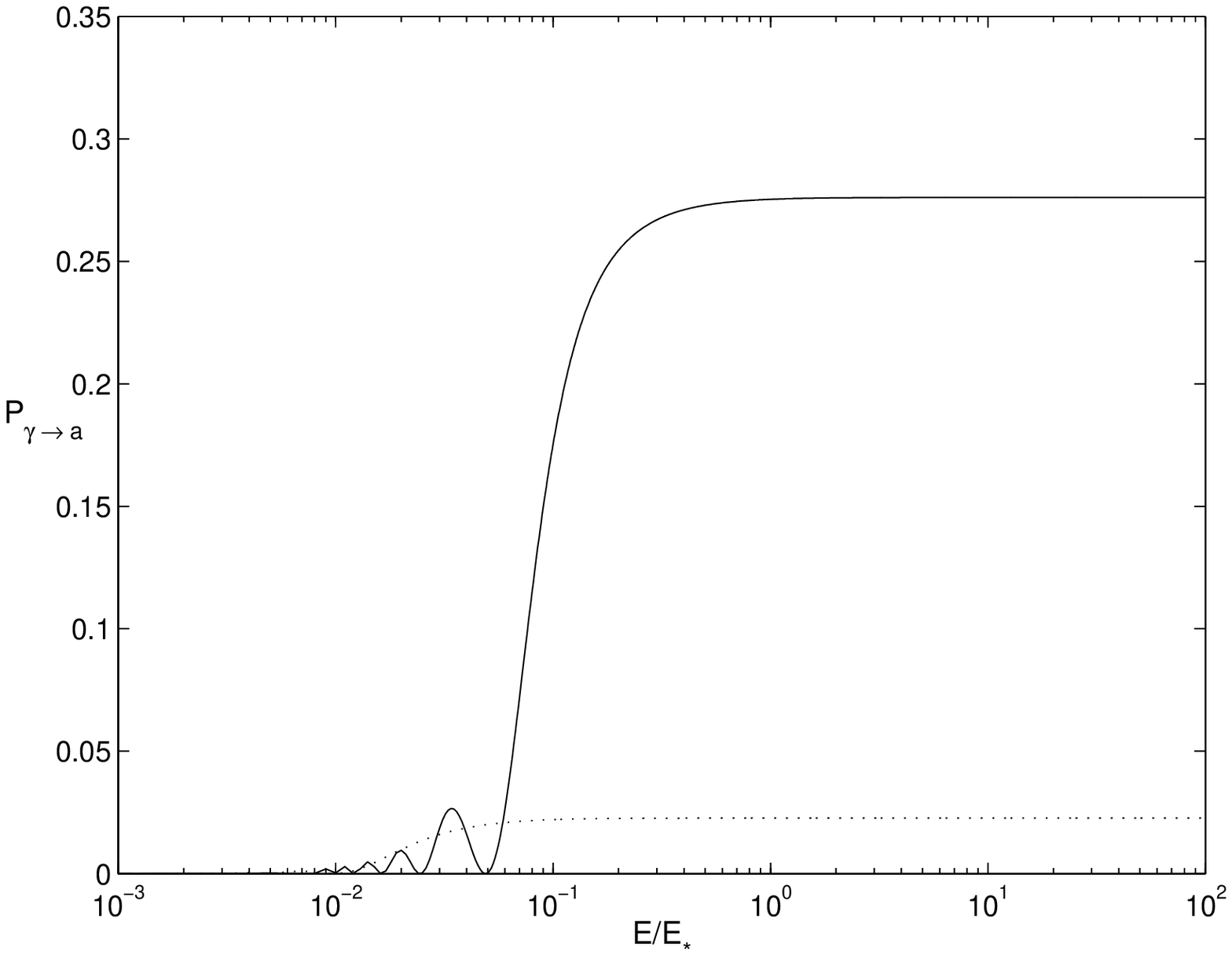}
\end{tabular}
\caption{\label{fig:P_omStar_extra_200}
Conversion probability versus photon energy in units of $E_*$ in the large-scale magnetic field.
The plots in the left panels show the conversion probability over a single magnetic domain $P_{\gamma \to \phi}^{(0)}$, whereas those in the right panels represent the total conversion probability $P_{\gamma \to \phi}$ over $N = 200$ magnetic domains.
The plots in the upper panels are obtained for $M = 3 \cdot 10^{10}$~GeV, while those in the lower panels arise for $M = 1 \cdot 10^{11}$~GeV.
Dotted and solid lines correspond to $B = 1 \cdot 10^{-9}$~G and $B = 5 \cdot 10^{-9}$~G, respectively.}
\end{figure}

\begin{figure}[t]
\begin{tabular}{@{}c@{}c@{}}                              \hspace{-5mm}
\includegraphics[width=.56\textwidth]{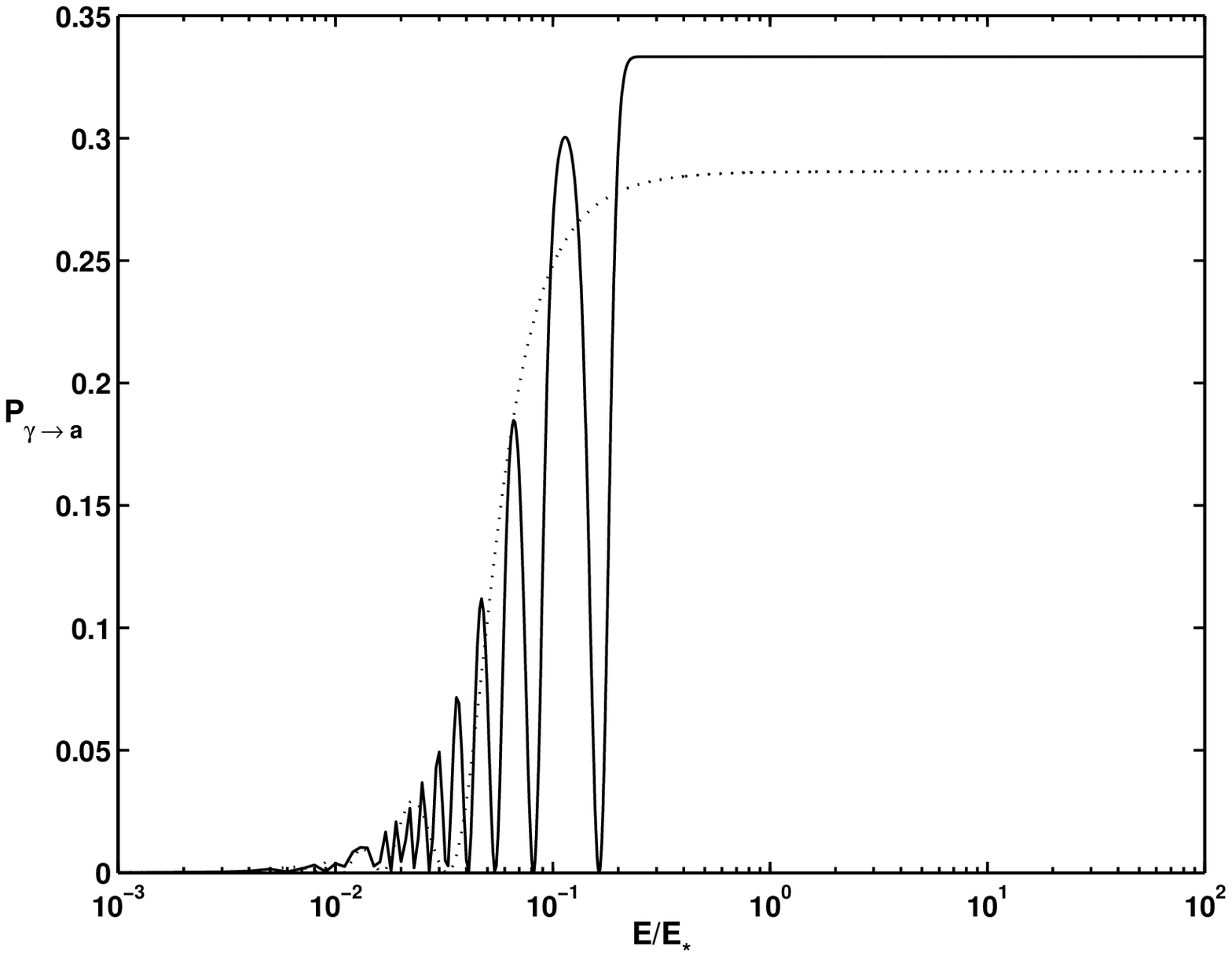} \hspace{-2.5mm}
&                                                         \hspace{-5mm}
\includegraphics[width=.56\textwidth]{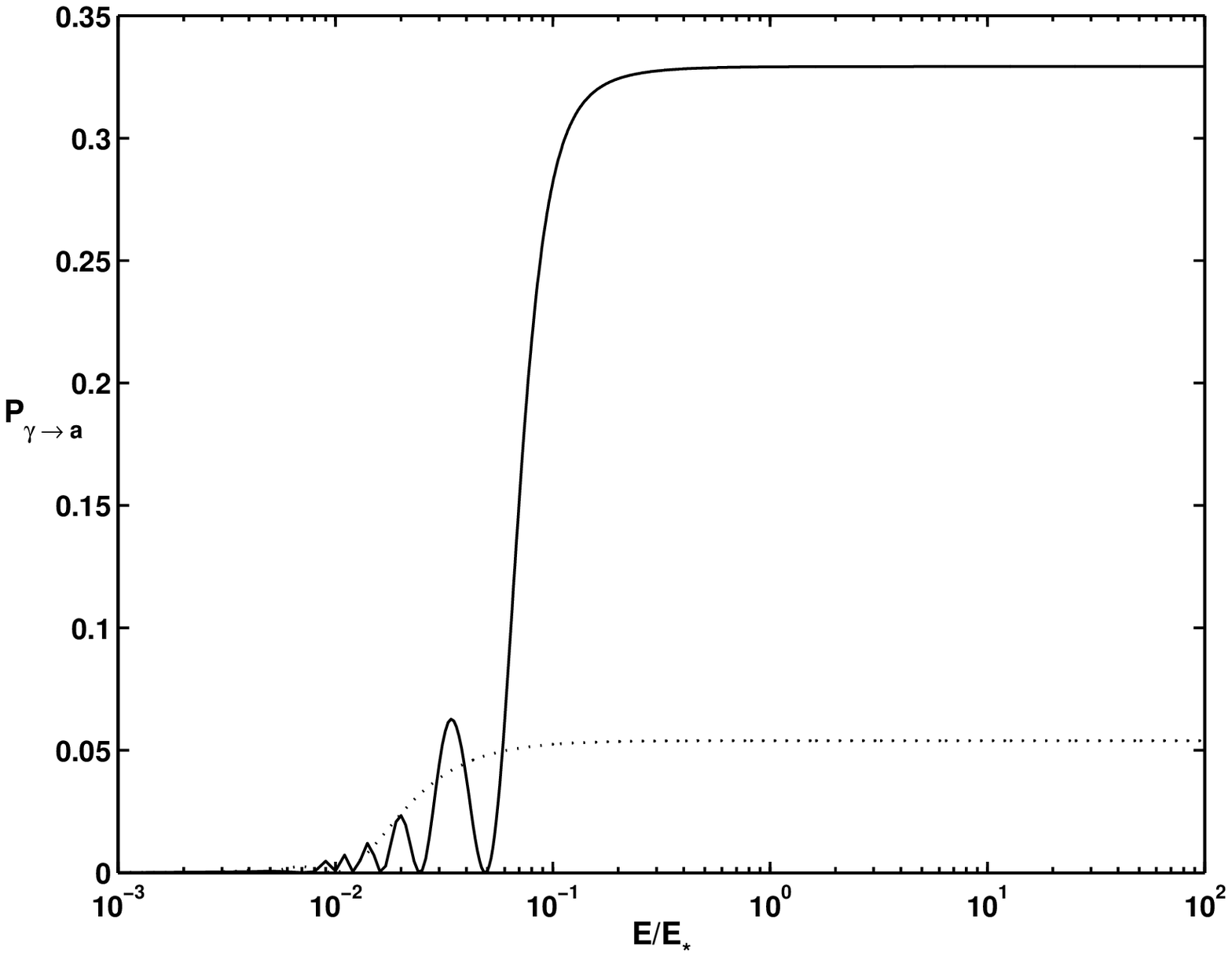}
\end{tabular}
\caption{\label{fig:P_omStar_extra_500}
Conversion probability $P_{\gamma \to \phi}$ versus photon energy in units of $E_*$ in the large-scale magnetic field: same as in the right panels of Fig.~\ref{fig:P_omStar_extra_200} but over $N = 500$ magnetic domains.
The left plot is obtained for $M = 3 \cdot 10^{10}$~GeV, while the right one arises for $M = 1 \cdot 10^{11}$~GeV.
Dotted and solid lines correspond to $B = 1 \cdot 10^{-9}$~G and $B = 5 \cdot 10^{-9}$~G, respectively.}
\end{figure}

The conversion probability in the large-scale magnetic field is shown in Fig.~\ref{fig:P_omStar_extra_200} and Fig.~\ref{fig:P_omStar_extra_500}.
We have computed the effect for source distances \mbox{$D=200$~Mpc} (Fig.~\ref{fig:P_omStar_extra_200}) and \mbox{$D=500$~Mpc} 
(Fig.~\ref{fig:P_omStar_extra_500}), which correspond to \mbox{$N=200$} and \mbox{$N=500$} magnetic domains crossed by the beam, respectively.
The left panels of Fig.~\ref{fig:P_omStar_extra_200} show the effect over a single domain of the magnetic field (this is obviously independent of distance).
We have considered two different values of the inverse photon coupling constant $M$ (\mbox{$M = 3 \cdot 10^{10}$~GeV} and \mbox{$M = 1 \cdot 10^{11}$~GeV}), and two different values of the magnetic field strength $B$ (\mbox{$B = 1 \cdot 10^{- 9}$~G} and \mbox{$B = 5 \cdot 10^{- 9}$~G}).

We see that for $M$ sufficiently close to the CAST lower bound (upper panels in Fig.~\ref{fig:P_omStar_extra_200} and left panel in Fig.~\ref{fig:P_omStar_extra_500}) the spectral distortion is observable provided $B$ is roughly within one order of magnitude from the upper limit.
A similar result shows up in the opposite situation, namely for $B$ sufficiently close to the upper bound (solid lines) and $M$ roughly within one order of magnitude from the CAST lower limit (lower panels in Fig.~\ref{fig:P_omStar_extra_200} and right panel in Fig.~\ref{fig:P_omStar_extra_500}).
This conclusion is practically unaffected by the source distance.

\subsection{Intracluster contribution}

\begin{figure}[h!t]
\begin{tabular}{@{}c@{}c@{}}                           \hspace{-5mm}
\includegraphics[width=.56\textwidth]{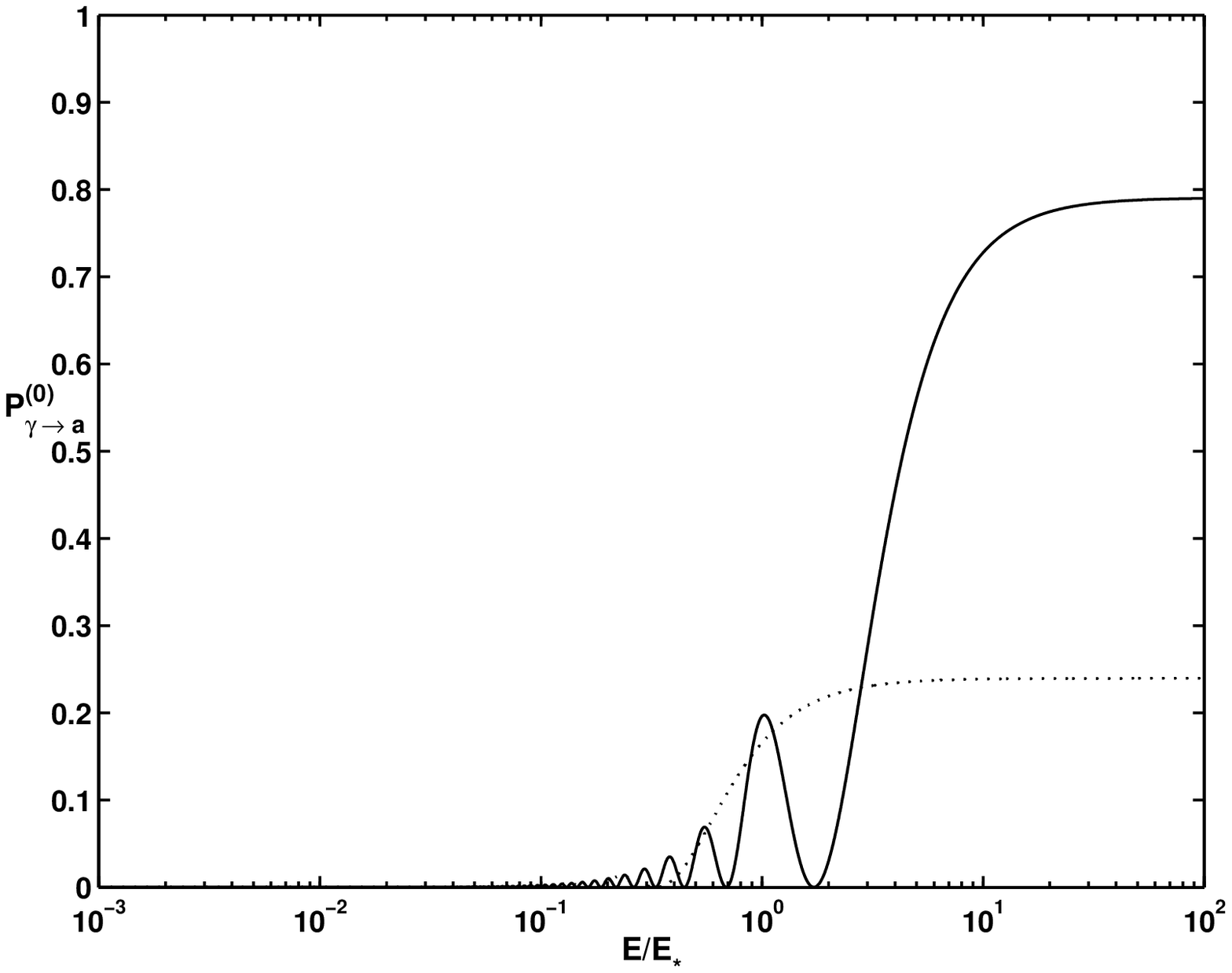} \hspace{-2.5mm}
&                                                      \hspace{-5mm}
\includegraphics[width=.56\textwidth]{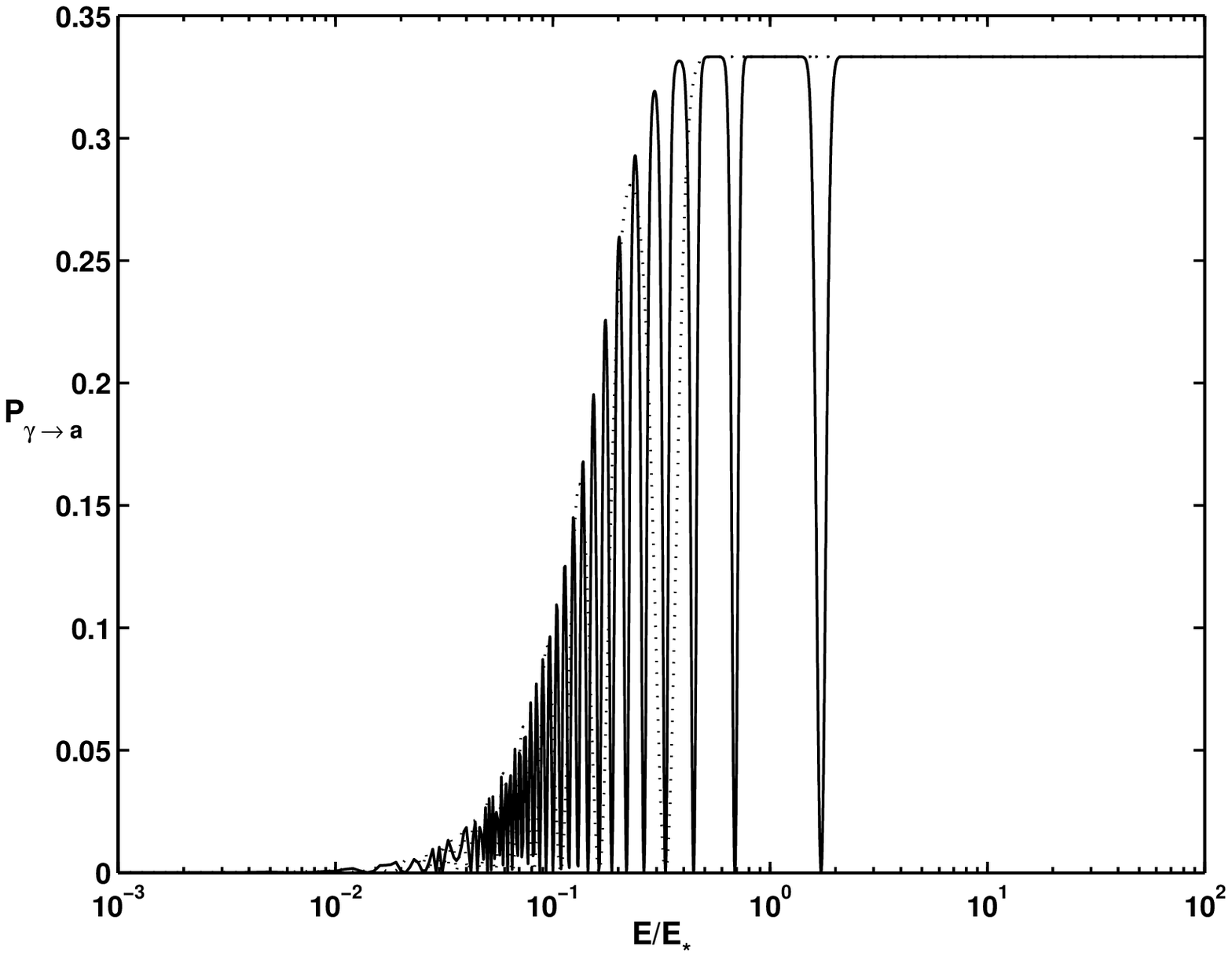}
\\                                                     \hspace{-5mm}
\includegraphics[width=.56\textwidth]{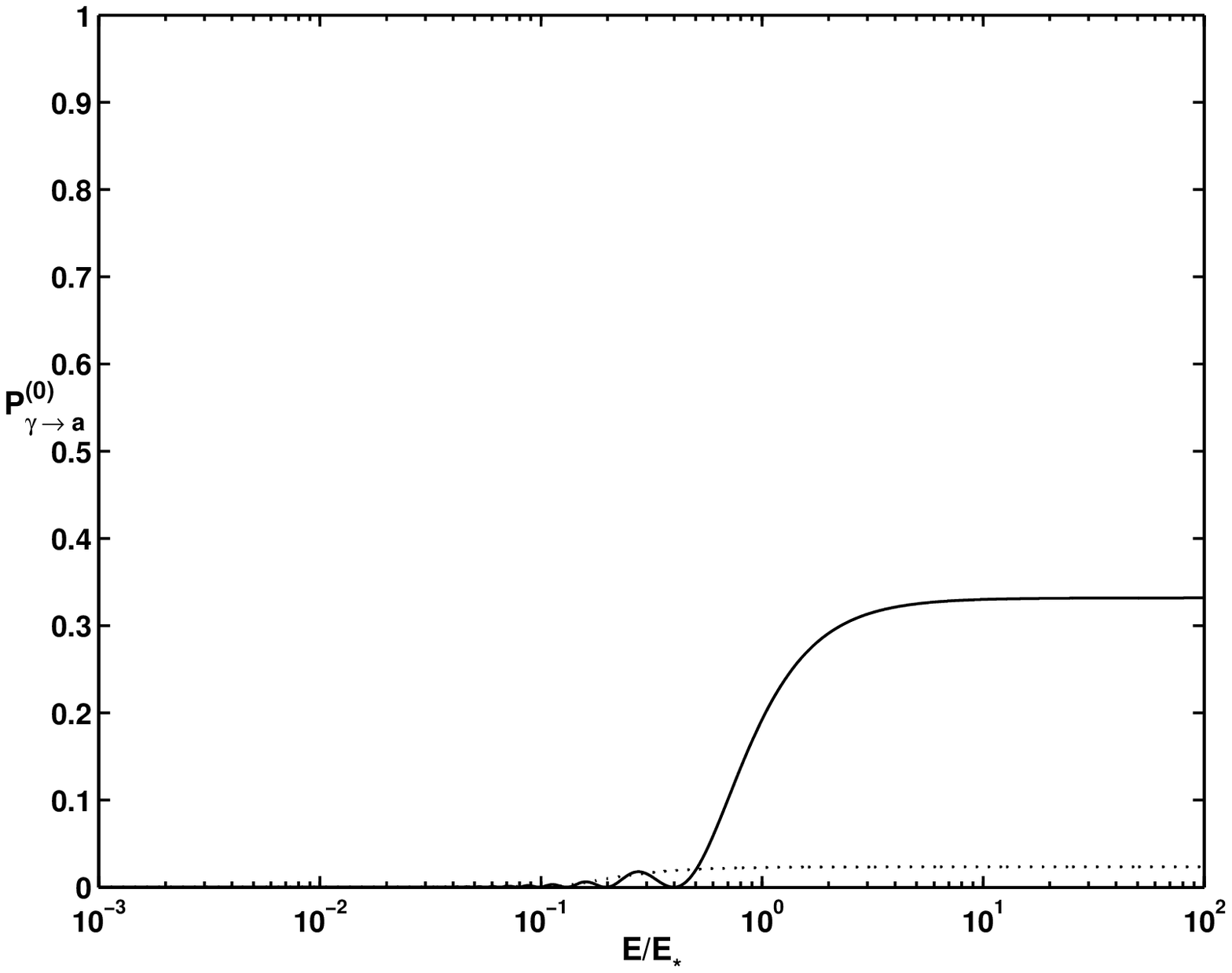} \hspace{-2.5mm}
&                                                      \hspace{-5mm}
\includegraphics[width=.56\textwidth]{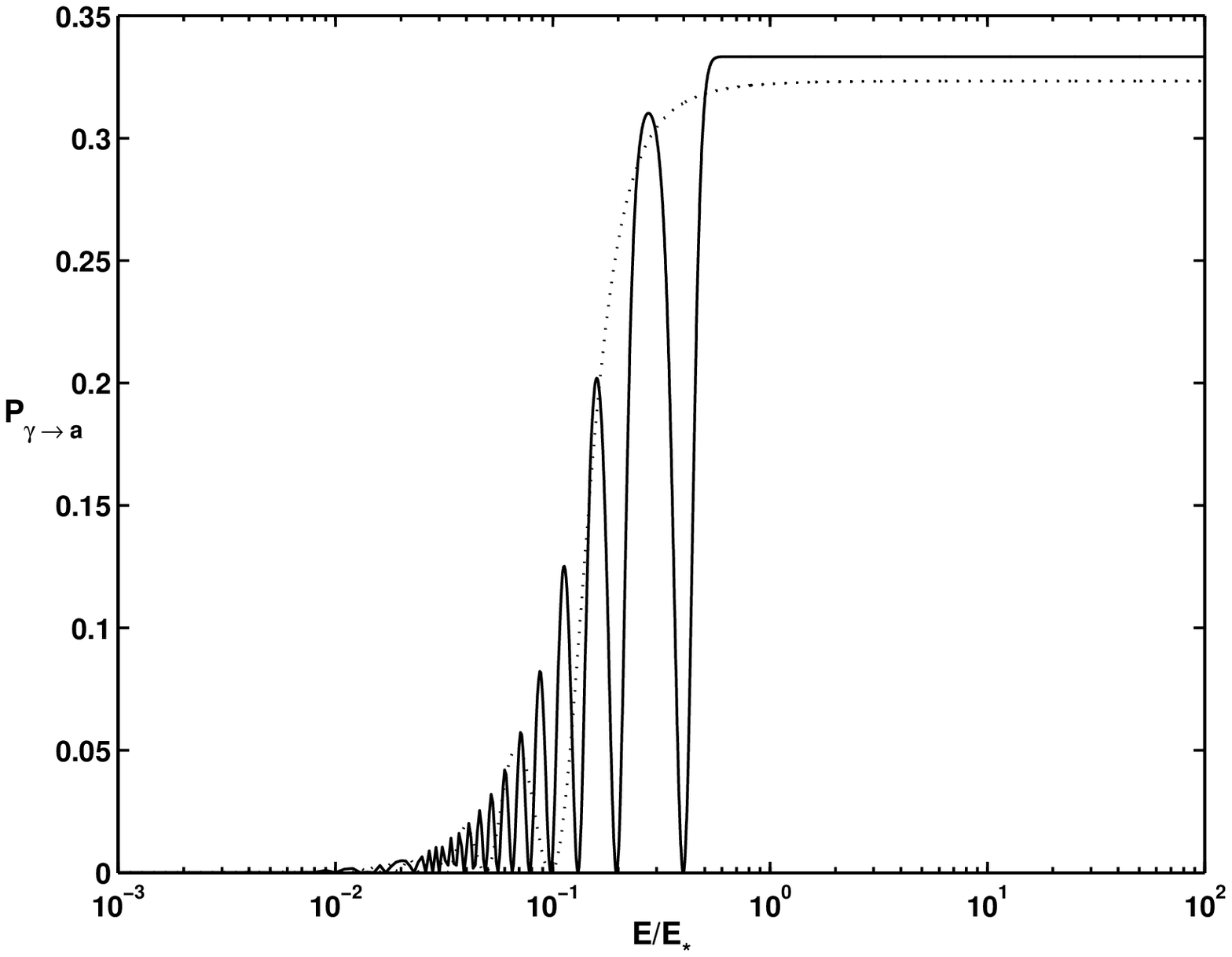}
\end{tabular}
\caption{\label{fig:P_omStar_intra}
Conversion probability versus photon energy in units of $E_*$ in intracluster magnetic fields as well as in the regular Galactic field.
The plots in the left panels show the conversion probability over a single magnetic domain $P_{\gamma \to \phi}^{(0)}$, whereas those in the right panels represent the total conversion probability $P_{\gamma \to \phi}$ over $N = 100$ magnetic domains.
The plots in the upper panels are obtained for $M = 3 \cdot 10^{10}$~GeV, while those in the lower panels arise for $M = 1 \cdot 10^{11}$~GeV.
Dotted and solid lines correspond to $B = 1 \cdot 10^{-6}$~G and $B = 4 \cdot 10^{-6}$~G, respectively.}
\end{figure}

The conversion probability in the magnetic field of a galaxy cluster is shown in Fig.~\ref{fig:P_omStar_intra}.
We assume that the cluster has a typical size of 1 Mpc, so that $N= 100$ magnetic domains are crossed by the beam.
These plots correspond to $M = 3 \cdot 10^{10}$~GeV and $M = 1 \cdot 10^{11}$~GeV, and to $B = 1 \cdot 10^{- 6}$~G and $B = 4 \cdot 10^{- 6}$~G.

Clearly, if the beam goes through a cluster of galaxies the spectral energy distortion turns out to be observable for all our preferred values of $M$ and $B$, that is to say provided these parameters lie roughly within one order of magnitude from their bounds.

\subsection{Galactic contribution}

The situation concerning photon-ALP conversion in the Milky Way can be summarized as follows.

{\it Regular component} --
The conversion probability in this magnetic field is computed directly from eq.~(\ref{a16x}) and it is shown in the left panels of Fig.~\ref{fig:P_omStar_intra}.
These plots again correspond to $M = 3 \cdot 10^{10}$~GeV and $M = 1 \cdot 10^{11}$~GeV, and to $B = 1 \cdot 10^{- 6}$~G and $B = 4 \cdot 10^{- 6}$~G.
The situation is analogous to what we found in the case of the large-scale magnetic field.
That is, for $M$ close enough to the CAST lower bound the spectral distortion is observable provided $B$ is roughly within one order of magnitude from the upper limit.
Similarly, observability is ensured for $B$ sufficiently close to the upper bound and $M$ roughly within one order of magnitude from the CAST lower limit.

{\it Turbulent component} --
It is straightforward to realize that we presently have $L_{\rm osc} \gg L_{\rm dom}$ for all experimentally allowed values of $m$ and $M$.
Therefore, \mbox{$P_{\gamma \to \phi}^{(0)}(L_{\rm dom}) \simeq (B_T \, L_{\rm dom}/(2 M))^2$} and we get \mbox{$P_{\gamma \to \phi}^{(0)} < 2.3 \cdot 10^{- 12}$} by enforcing \mbox{$M > 10^{10} \, {\rm GeV}$}.
Because here the number of magnetic domains is $N \sim 10^6$, we end up with \mbox{$P_{\gamma \to \phi} < 10^{- 5}$}.
Thus, we see that in this case no spectral energy distortion is observable.

\section{Discussion and Conclusions}

We have presented a mechanism whereby the energy spectra of gamma-ray sources at cosmological distances get distorted due to photon-ALP mixing taking place in the cosmic magnetic fields crossed by the beam on its way to us.
We have attempted to identify the ranges of the ALP parameters for which the effect in question can be observed with GLAST.
Unfortunately, the uncertainties in the properties of cosmic magnetic fields prevent us from making sharp statements about an exclusion plot in the ALP parameter space.
Nevertheless, we have succeeded in showing that observability is achieved for ranges of the ALP parameters which are allowed by all available constraints.

Large-scale magnetic fields as well as the regular Galactic component turn out to be nearly equally efficient at producing an observable distortion.
Whenever the beam crosses a cluster of galaxy, its intracluster magnetic field is even more efficient in that respect.
The latter circumstance suggests to look at similar sources however in different directions, so that cluster crossing occurs only for one line of sight.
Directionality can also be instrumental in detecting the spectral enery distortion due to the Galactic regular magnetic field, since its morphology is presently fairly well known.

Our proposal shares some similarities with the one advanced a few years ago by Csaki, Kaloper and Terning (CKT)~\cite{Csaki} as an explanation for the observed dimming of distant type Ia supernovae~\cite{supernovae}.
Currently, such a dimming is interpreted as evidence for an accelerated cosmic expansion, presumably triggered by a mysterious dark energy~\cite{cosmology}.
Instead, CKT suggested that the supernovae under consideration look fainter than expected simply because some photons en route to us become ALPs in extragalactic magnetic fields, thereby escaping detection.
Unfortunately, subsequent studies have shown that this proposal gets ruled out for almost all values of the parameter space~\cite{mirizzi1}.
In particular, plasma effects make the dimming of type Ia supernovae excessively chromatic.
We stress that these problems are {\it automatically} avoided in our case simply because any dimming effect disappears at energy
$E \ll E_* \sim 10^2 \, {\rm GeV} - 1 \, {\rm TeV}$.

A somewhat different idea has recently been put forward by Hooper and Serpico~\cite{serpico}, and by Hochmuth and Sigl~\cite{sigl}.
Here, it is proposed that photon-ALP conversion can take place {\it inside} gamma-ray sources, thanks to their strong magnetic fields.
We have seen in Sect.~2 that efficient photon-ALP conversion requires the strong-mixing regime to be realized.
Moreover, it follows from eq.~(\ref{a16}) that $P_{\gamma \to \phi}^{(0)}(x)$ becomes maximal for $\Delta_{\rm osc} \, x \sim 1$, namely for $B_T/M \, x \sim 1$ (this just follows from eq.~(\ref{a17}) in the strong-mixing case).
The latter condition is similar the Hillas criterion~\cite{hillas} concerning the acceleration of cosmic-ray particles, and this circumstance is used to argue that such a regime should take place in some astrophysical sources.
Manifestly, both this mechanism and the one proposed in this Letter can be operative at the same time.

We thank Sasha Dolgov and Dario Grasso for discussions.

\end{document}